\documentclass{article}
\usepackage{spconf,amsmath,graphicx,hyperref}
\usepackage{booktabs}
\usepackage{dingbat}
\usepackage{fvextra}
\fvinlineset{breaklines,breakanywhere,breakanywheresymbolpre=}

\title{JSIR: An Audio-Aligned, Multi-Task Dataset of Jazz Standard Recordings}
\name{Sihun Lee\thanks{Corresponding author: sihunlee@sogang.ac.kr}, Dasaem Jeong}
\address{Sogang University\\
 Department of Artificial Intelligence\\
 Seoul, South Korea}
\begin{document}
\ninept
\maketitle
\begin{abstract}
Beat tracking, chord estimation, and structure analysis are staple tasks of music information retrieval. Their datasets are drawn largely from Western popular music, and jazz remains a repertoire on which state-of-the-art systems underperform. We present JSIR, the Jazz Standards dataset for Information Retrieval. Its 267 recordings of 44 jazz standards span 25.6 hours, 3.7 times the audio of JAAH, the closest existing resource. For each recording, we provide annotations for beat and downbeat positions, chords, and sections, all produced by a jazz musician and revised in several passes.
In baseline experiments, models trained with JSIR also improve on other jazz datasets: the downbeat F-measure of a state-of-the-art beat tracker rises from 87.7\% to 89.2\%, and the seventh-chord accuracy of a pop-trained chord estimator on JAAH from 34\% to 41\%.
We publicly release the annotations, the annotation tool used to create them, and a web application that displays them in sync with the audio.
\end{abstract}
\section{Introduction}
\label{sec:intro}

Beat tracking, automatic chord estimation (ACE), and music structure analysis are staple tasks of music information retrieval (MIR), and progress on each has been driven by audio recordings paired with time-aligned annotations. Much of the datasets that shaped these tasks are drawn from Western popular music, and systems trained on them, from Bayesian networks~\cite{krebs2015statespace} and convolutional chord recognizers~\cite{korzeniowski2016fullyconv} to transformer models~\cite{foscarin2024beatthis,kim2023allinone}, inherit the assumptions of that repertoire: a drum kit that marks the beat, a chord vocabulary dominated by diatonic triads, and a verse--chorus form in which repetition signals structure.

Jazz violates each of these assumptions. Rhythmically, the swing feel places the pulse in the ride cymbal and the walking bass rather than in a kick-and-snare pattern, tempo is negotiated rather than fixed, and soloists phrase freely across the beat. Harmonically, seventh chords and their extensions are the norm. Piano or guitar comping states the harmony sparsely and irregularly over an improvised bass line, so the onset of a chord is often ill-defined~\cite{eremenko2018jaah}. Formally, a performance typically follows the head--solo--head schema: the ensemble states the theme, soloists improvise in turn over repeated cycles (choruses) of the same harmonic form, and the theme returns. Since every chorus repeats the same harmony, the repetition cues on which structure analysis relies carry no boundary information, and changes in sections are instead delineated by changes of soloist and instrumentation~\cite{balke2022jsd}. The consequences are measurable: a state-of-the-art beat tracker loses more than 12 percentage points of F-measure on RWC-Jazz relative to RWC-Popular~\cite{foscarin2024beatthis}, a strong pop-trained chord recognizer reaches only 40\% accuracy on a five-class jazz vocabulary~\cite{eremenko2018jaah}, and a SALAMI-trained boundary detector halves its F-measure on jazz recordings~\cite{balke2022jsd}.

Progress is limited less by modeling ideas than by data. Existing jazz resources are valuable but partial. The Weimar Jazz Database~\cite{pfleiderer2017jazzomat} provides symbolic transcriptions of solos rather than annotations of complete recordings; the Jazz Structure Dataset~\cite{balke2022jsd} annotates structure alone; and the Jazz Audio-Aligned Harmony dataset (JAAH)~\cite{eremenko2018jaah}, the closest precedent to our work, only covers a single performance for most of the listed compositions.

In this context, we present JSIR, the Jazz Standards dataset for Information Retrieval. It consists of 267 commercially released recordings of 44 jazz standards, about 29 hours of audio in total. Each recording is annotated with beat and downbeat positions, chord labels, and section labels with instrument tags. Two principles guide its design. It is \emph{multi-task}: the three annotation layers share one beat grid, so every chord and section boundary falls on a beat. It is \emph{multi-version}: each standard is represented by four to nine recordings, which opens jazz to the cross-version methods that multi-performance corpora have enabled in classical music~\cite{konz2013crossversion,weiss2021swd}. 
We publicly release the annotations along with a web application for browsing the dataset's contents.\footnote{\url{https://github.com/MALerLab/JSIR}}

\begin{figure}[t]
  \centering
  \includegraphics[width=0.95\linewidth]{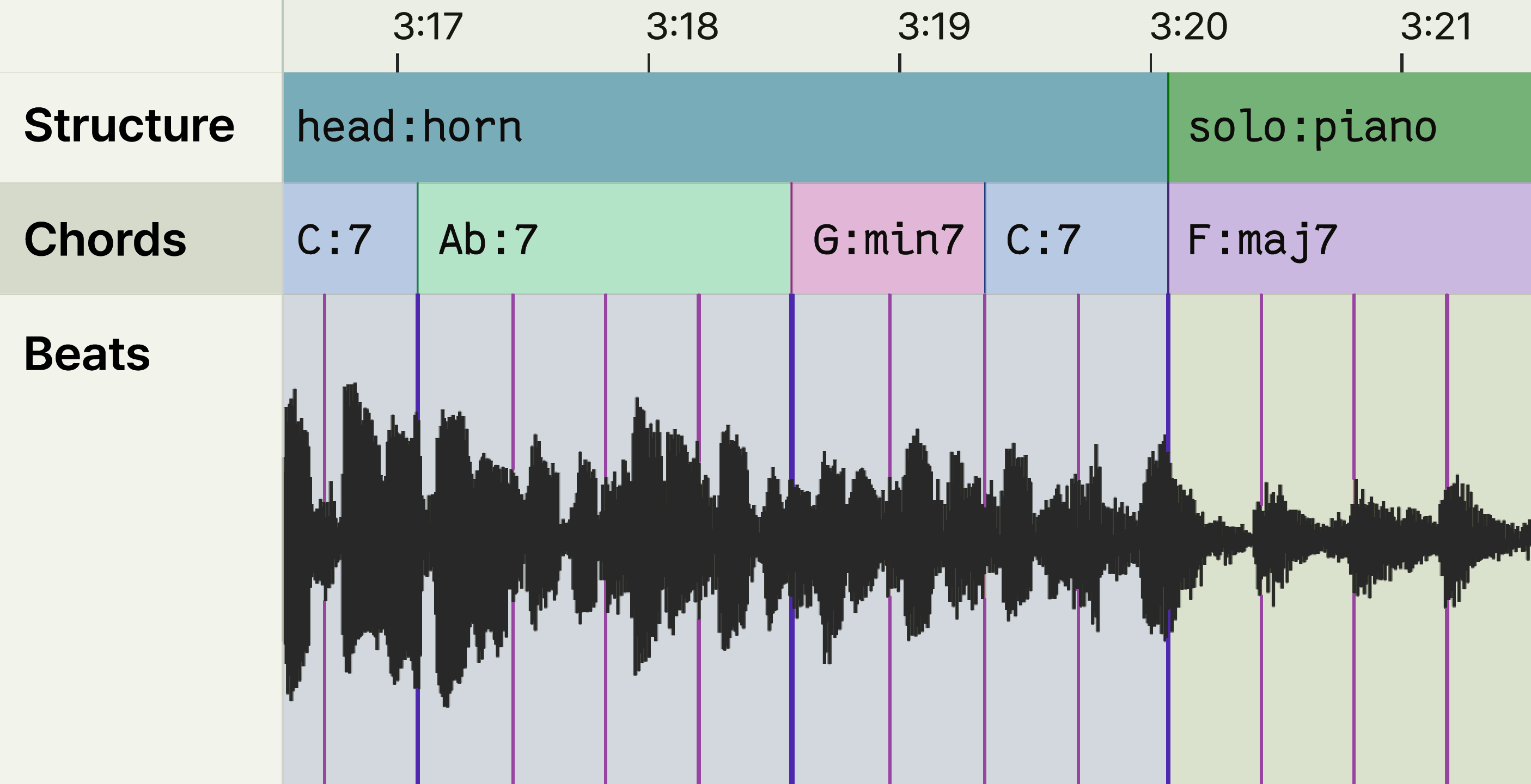}
  \caption{An example illustration of the annotations in JSIR. The chord and structure labels are aligned to the beat positions.}
  \label{fig:main}
\end{figure}

\section{Related Work}
\label{sec:related}

\subsection{Annotated Datasets for Rhythm, Harmony, and Structure}

Annotated datasets have driven research in various tasks in MIR: Namely, Ballroom~\cite{gouyon2006ballroom} and GTZAN~\cite{marchand2015gtzan} in beat tracking; Isophonics~\cite{mauch2009omras2} and Billboard~\cite{burgoyne2011billboard} in ACE; SALAMI~\cite{smith2011salami} and Harmonix~\cite{nieto2019harmonix} in structure analysis. Only a few corpora annotate several layers of the same audio: Isophonics provides chords, keys, beats, and sections for 179 Beatles songs~\cite{mauch2009omras2}, and Harmonix provides joint beat, downbeat, and section labels that enabled multi-task models such as All-In-One~\cite{kim2023allinone}.

Jazz representation is marginal in such resources. Where in-domain data have been supplied for other under-represented styles, the gains have been large: less than a minute of annotated candombe~\cite{maia2022latinamerican} and fine-tuning on Carnatic music~\cite{nadar2025carnatic} each transformed off-the-shelf performance. A comparable resource is what jazz has lacked.

\subsection{Jazz Resources}

The Weimar Jazz Database (WJD)~\cite{pfleiderer2017jazzomat}, the outcome of the Jazzomat project, is the principal earlier resource for computational jazz research: it contains monophonic solo transcriptions along with beat and chord annotations, but for excerpts rather than for entire recordings.
The Jazz Structure Dataset (JSD)~\cite{balke2022jsd} annotates the complete WJD recordings at the chorus and solo levels with instrument labels; it carries no chord or beat annotations for the full tracks. JAAH~\cite{eremenko2018jaah} annotates 113 tracks with beats produced by the madmom tracker and corrected by hand, lead-sheet-style chords per bar, and a nested structure with chorus form and instrumentation.
Filosax~\cite{foster2021filosax} contains about 24 hours of tenor saxophone recorded by five players over 48 commercial play-along tracks, but the pre-recorded rhythm section cannot interact with the soloist.
JSIR complements these resources: it carries all three annotation layers on one beat grid for complete recordings, and it contains several versions of every jazz standard.

\subsection{Jazz as a Challenge for MIR Systems}

The literature repeatedly treats jazz as an outlier rather than a target: it typically appears as one held-out row in an evaluation table, and the gap has persisted through two decades of architectural change~\cite{durand2017ensemble,foscarin2024beatthis,eremenko2018jaah,balke2022jsd}. The findings recur across studies. 
Durand et al.~\cite{durand2017ensemble} report frequent metrical-level errors in downbeat tracking on RWC-Jazz and note that instruments may play rhythmically freely, especially during solos.
Chord recognizers are trained and evaluated on vocabularies centered on triads, so the seventh-chord harmony of jazz falls into a poorly modeled tail, while the sparse, syncopated nature of comping leaves chord onsets ill-defined; a jazz-specific system sidesteps both problems by aligning and pooling repeated choruses~\cite{duran2020transcribing}. Previous methods in structure analysis build on repetition, homogeneity, and novelty~\cite{nieto2020survey}. In jazz, however, section boundaries are marked by timbral change rather than by repetition, and learned representations capture this cue only in part~\cite{buisson2024taslp}. Finally, multi-version corpora have enabled cross-version analysis and version identification in other repertoires~\cite{konz2013crossversion,weiss2021swd,yesiler2021version}; jazz standards, whose shared form underlies radically different arrangements, are a natural testbed that few existing audio datasets support.

\section{Dataset Construction}
\label{sec:method}
As illustrated in Figure~\ref{fig:main}, JSIR contains a comprehensive set of annotations for each recording. All annotations were produced by the author (a jazz musician) using a purpose-built tool and thoroughly revised in several passes. The following sections describe each step in detail.

\subsection{Repertoire and Recording Selection}

JSIR is organized around jazz standards, the compositions that jazz musicians share as a common repertoire. We selected 44 standards spanning the Great American Songbook, the blues, and bebop and hard-bop originals (e.g., \emph{All the Things You Are}, \emph{Bags' Groove}, \emph{Confirmation}, \emph{Giant Steps}). For each standard we encoded the chord progression from the canonical lead sheet in Harte syntax~\cite{harte2005symbolic} with one chord slot per beat, together with its key, time signature, tempo class, and rhythm feel, following the WJD classifications. 

Candidate recordings were enumerated from the standard's MusicBrainz\footnote{\url{https://musicbrainz.org}} work entry, ranked by the number of releases as a proxy for prominence, and screened by ear. Between four and nine recordings were kept per standard, 267 in total, their instrumentation ranging from small combos to vocal performances with orchestra and big bands. Because the audio is under copyright, we distribute the MusicBrainz and YouTube identifiers of the annotated audio, so that users can retrieve the exact version and hence the exact timing.

\subsection{Beat and Downbeat Annotation}

Initial beat estimates were produced with the beat tracker of the madmom library~\cite{bock2016madmom}, which decodes a recurrent-network activation with a bar-pointer dynamic Bayesian network~\cite{krebs2015statespace}. The estimates are then corrected by hand. Typical interventions were octave errors (tracking at half or double tempo, common in fast bebop and ballads), phase errors (locking onto the off-beat under strong swing), and drift in passages with a weak or absent pulse, such as rubato introductions, final ritardandi, and drum solos, where beats were placed manually. Downbeats were then marked by hand: after the first downbeat of each chorus is located and the following bars gridded accordingly, the beat grid was corrected for irregular bars such as added or dropped beats in breaks and metric changes in codas.

\subsection{Chord Annotation}

Chord labels were initialized from lead sheet progressions; for each standard, we compared lead sheets from several sources and assembled a progression that the largest number of recordings agree on. For each head and solo section, the progression was transposed to the key of the recording and superimposed on the beat grid so that each chord slot coincides with the corresponding beat. The labels were then revised against the audio in three ways. First, chords were corrected where the performance departs from the lead sheet, for instance through substitute turnarounds, tritone substitutions, or the alternative changes carried by a particular arrangement. Second, chord events were removed, and the no-chord label \texttt{N} inserted, where the audio carries no harmonic information, as in unaccompanied drum and bass solos, breaks, and free introductions or endings. Third, new chords were added for material absent from the lead sheet, such as arranged interludes, vamps, and codas. Labels follow the Harte syntax~\cite{harte2005symbolic} in its shorthand form (e.g., \texttt{F:min7}, \texttt{G:hdim7}, \texttt{C:7(b9)}); following JAAH, they are to be read as jazz chord symbols rather than literal pitch-class sets, and inversions are given only for composed bass lines.

\subsection{Structure Annotation}

Structure was annotated manually with a six-label taxonomy tailored to the head--solo--head schema: \texttt{intro}, \texttt{head}, \texttt{solo}, \texttt{trade}, \texttt{interlude}, and \texttt{outro}. A \texttt{head} is a statement of the theme; a \texttt{solo} is a stretch of improvisation by one soloist, possibly spanning several choruses; a \texttt{trade} is a passage in which two or more soloists alternate within a chorus (colloquially, \textit{trading fours or eights}); an \texttt{interlude} is arranged material between sections; and \texttt{intro} and \texttt{outro} denote material before the first and after the last statement of the form.

For \texttt{head}, \texttt{solo}, and \texttt{trade}, the prominent instruments are appended after a colon and separated by commas when several are involved, e.g., \Verb{head:clarinet, solo:piano, trade:horn,drum}, which supports solo detection and instrument recognition. Section boundaries are placed on the downbeat that begins the harmonic form, so that pickups belong to the preceding section, in line with JSD.

\subsection{Data Format and Scale}

Each recording is described by a metadata record (standard, artist, album, instrumentation, key, time signature, bars per chorus, tempo class, rhythm feel, and the MusicBrainz and YouTube identifiers) and four plain-text annotation files: beats (time and downbeat flag), chords (onset time and label), structure (onset time and label), and \emph{segments}, the time spans in which both the beat grid and the harmonic annotation are reliable enough for training and evaluation, which exclude rubato introductions and endings, solos with ambiguous timing, and fade-outs. The annotations are released together with a web application that renders them synchronously with the audio for inspection and listening.

Table \ref{tab:stats} compares JSIR's composition and scale against other jazz datasets. JSIR differs from JAAH, the closest existing resource, in two ways: It is 3.7 times larger in audio duration, with 267 recordings against 113; and every standard appears in four to nine versions, where JAAH has one for most. Like JAAH, and unlike WJD and JSD, it carries all three layers on one beat grid for complete recordings.

% \section{Dataset Statistics}
% \label{sec:stats}

\begin{table}[t]
    \centering
    \small
    \begin{tabular}{l|ccccr} \toprule
         Dataset&  Beats&  Chords&  Structure&  Songs& Duration\\ \midrule
         WJD~\cite{pfleiderer2017jazzomat}&  \checkmark&  \checkmark&  &  344& 12.8 h\\
         JSD~\cite{balke2022jsd}&  &  &  \checkmark&  340& 33.4 h\\
         JAAH~\cite{eremenko2018jaah}&  \checkmark&  \checkmark&  \checkmark&  113& 6.9 h\\
         JSIR&  \checkmark&  \checkmark&  \checkmark&  267& 25.6 h\\ \bottomrule
    \end{tabular}
    \caption{Comparison between JSIR and three other jazz-focused datasets: three label types, number of recordings, and the duration of audio covered (in hours).}
    \label{tab:stats}
\end{table}

% corpus-level statistics (recordings, standards, duration, versions per
% standard, instrumentation, tempo/feel, chord vocabulary, structure labels).

\section{Experimental Results}
\label{sec:results}

% NOTE (beat tracking): Table (a) and Sec. 5.1 use the 2026-09-24 run on the
% final release (267 recordings, 340 segments, 20.6 h); see
% beat_this/repro/results/jsir_final_cv_comparison.txt and PROGRESS.md, Sec. 13.6.
% NOTE (structure): 12 of the 267 recordings were excluded from the structure
% experiments (6 empty structure files, 5 with placeholder labels, 1 audio file
% with 33 min of un-annotated material); see SongFormer/JSIR_EXPERIMENTS.md, Sec. 2.

\begin{table*}[t]
  \centering
  \small
  \setlength{\tabcolsep}{5pt}
  \begin{tabular}{@{}l ccc ccc cc@{}}
    \toprule
    \multicolumn{9}{@{}l}{\textbf{(a) Beat and downbeat tracking} --- Beat This!, 8-fold cross-validation; F-measure and CMLt in \%} \\
    \midrule
    & \multicolumn{3}{c}{Jazz test sets (211 tracks)} & \multicolumn{3}{c}{JSIR (267 recordings)} & \multicolumn{2}{c}{Non-jazz datasets} \\
    \cmidrule(lr){2-4} \cmidrule(lr){5-7} \cmidrule(l){8-9}
    Training data & Beat F & Downbeat F & Downbeat CMLt & Beat F & Downbeat F & Downbeat CMLt & Beat F & Downbeat F \\
    \midrule
    16 datasets~\cite{foscarin2024beatthis} & 93.4 & 87.7 & 79.0 & 94.6 & 86.5 & 72.7 & \textbf{89.0} & \textbf{86.2} \\
    16 datasets + JSIR & \textbf{93.6} & \textbf{89.2} & \textbf{81.4} & \textbf{97.1} & \textbf{90.6} & \textbf{81.1} & 88.9 & 86.0 \\
    \bottomrule
  \end{tabular}

  \vspace{10pt}
  \begin{tabular}{@{}l cccc cccc@{}}
    \toprule
    \multicolumn{9}{@{}l}{\textbf{(b) Chord estimation} --- Consonance-ACE; duration-weighted \texttt{mir\_eval} scores in \%} \\
    \midrule
    & \multicolumn{4}{c}{JAAH (102 tracks)} & \multicolumn{4}{c}{JSIR (264 recordings)} \\
    \cmidrule(lr){2-5} \cmidrule(l){6-9}
    Training data & Root & MajMin & Sevenths & MIREX & Root & MajMin & Sevenths & MIREX \\
    \midrule
    Pop~\cite{poltronieri2025discord}, zero-shot & 54.4 & 48.3 & 33.7 & 55.6 & 49.6 & 38.8 & 23.1 & 50.8 \\
    JSIR, from scratch (5-fold CV) & 59.3 & 53.9 & 36.8 & 61.0 & 65.3 & 58.4 & 48.3 & 67.1 \\
    Pop $\rightarrow$ JSIR, fine-tuned (5-fold CV) & \textbf{62.7} & \textbf{58.9} & \textbf{41.3} & \textbf{65.1} & \textbf{65.8} & \textbf{59.5} & \textbf{48.9} & \textbf{68.5} \\
    \midrule
    \multicolumn{9}{@{}l}{\emph{Training-set size, from scratch, single hold-out split (JSIR columns: 47 held-out recordings$^\dagger$)}} \\
    \quad 25\% (54 recordings, 4.2~h) & 54.5 & 48.2 & 32.6 & 56.3 & 63.6 & 57.1 & 46.2 & 67.0 \\
    \quad 50\% (108 recordings, 8.4~h) & 58.2 & 53.4 & 36.9 & 60.1 & 66.1 & 61.3 & 48.8 & 70.5 \\
    \quad 100\% (217 recordings, 17.1~h) & 60.1 & 55.9 & 39.2 & 61.9 & 69.9 & 64.5 & 53.4 & 71.2 \\
    \bottomrule
  \end{tabular}

  \vspace{10pt}
  \begin{tabular}{@{}l l ccccc@{}}
    \toprule
    \multicolumn{7}{@{}l}{\textbf{(c) Structure analysis} --- SongFormer, 5-fold cross-validation, 3 seeds; scores in \%} \\
    \midrule
    Training data & Post-processing & HR.5F & HR.5F$_\mathrm{t}$ & HR3F$_\mathrm{t}$ & PWF & Accuracy \\
    \midrule
    Majority label (\texttt{solo}) & --- & 49.4 & 0.0 & 0.0 & 68.9 & 50.8 \\
    JSIR & default & 56.1 & 44.4 & 47.2 & \textbf{85.8} & \textbf{85.4} \\
    Pop checkpoint~\cite{hao2025songformer} $\rightarrow$ JSIR & default & 59.0 & 47.0 & 50.7 & 85.2 & 84.7 \\
    JSIR & tuned on validation & \textbf{79.0} & \textbf{68.2} & \textbf{70.2} & 85.1 & 83.9 \\
    Pop checkpoint $\rightarrow$ JSIR & tuned on validation & 78.5 & 67.1 & 69.3 & 83.7 & 82.3 \\
    \bottomrule
  \end{tabular}
  \caption{Baseline results for the three tasks. (a) Held-out jazz test sets are RWC-Jazz, JAAH, and Filosax; on JSIR the first row is zero-shot; non-jazz: the remaining training sets of Beat This! (4345 tracks, 3533 with downbeat annotations). (b) Cross-validated rows are means over the five fold models on JAAH and pooled held-out folds on JSIR; three overlapping tracks were excluded. $^\dagger$The 47 validation recordings (eight standards) of the hold-out split, on which the pop model reaches a root accuracy of 59.2. (c) HR: boundary hit-rate F-measure at 0.5\,s and 3\,s, subscript t: interior boundaries only; PWF: pairwise frame clustering; Accuracy: frame-level label accuracy. Bold marks the best value per column; the training-set-size rows use a different test subset and are left unmarked.}
  \label{tab:results}
\end{table*}

To establish baselines for the three annotation layers, we trained one recent, publicly available model per task on JSIR: Beat This!~\cite{foscarin2024beatthis} for beat and downbeat tracking, Consonance-ACE~\cite{poltronieri2025discord} for chord estimation, and SongFormer~\cite{hao2025songformer} for structure analysis. Architectures and training recipes follow the released code; only the data changed. Because different recordings of one standard share form and changes, all splits group whole standards, so that no tune occurs in both training and test. Table~\ref{tab:results} summarizes the results.

\subsection{Beat and Downbeat Tracking}

We followed the eight-fold cross-validation protocol of Beat This! and trained two models with identical code, seeds, and folds: one on the sixteen datasets of the original paper, one on the same data plus the reliable segments of JSIR (340 segments, 20.6~h), with matched optimizer steps. JSIR standards that share a tune or a recording with an external test track were pinned to that track's fold. Each held-out fold was scored on the jazz sets RWC-Jazz~\cite{goto2002rwc}, JAAH, and Filosax (211 tracks in total), on JSIR itself, and on the non-jazz datasets.

Adding JSIR leaves the beat F-measure unchanged everywhere but raises the downbeat F-measure on the external jazz sets by 1.5 percentage points (95\% CI 0.5--2.6; fold-level paired $t$-test, $p=0.03$). The gain comes from JAAH (+3.0 points); Filosax, whose play-along rhythm section is already tracked almost perfectly, and RWC-Jazz do not move. Continuity metrics gain more (+2.4 points for downbeat CMLt and AMLt): the model holds the bar phase over longer stretches. On JSIR, where the baseline is zero-shot, downbeat F-measure rises by 4.1 points and CMLt by 8.4. On the non-jazz datasets both F-measures stay within a pre-specified non-inferiority margin of 0.5 points, although the downbeat F-measure slips by 0.2 points and its confidence interval reaches the margin. The pulse of jazz is thus already tracked reliably; the bar level is the bottleneck, and a modest amount of in-domain data relieves it at little or no cost elsewhere.

\subsection{Chord Estimation}

The released Consonance-ACE checkpoint was trained on Isophonics and Billboard. We compared it applied zero-shot, the architecture trained from scratch on JSIR, and the checkpoint fine-tuned on JSIR. Training used the segments of 264 recordings (20.4~h) with pitch-shift augmentation over one octave; three JSIR recordings that are audio-identical to JAAH tracks were excluded throughout. Models were evaluated in-domain by five-fold cross-validation and out-of-domain on JAAH, with duration-weighted \texttt{mir\_eval}~\cite{raffel2014mir_eval} scores.

The pop-trained model attains a root accuracy of 54\% on JAAH and 50\% on JSIR and recognizes barely a third of the seventh chords. Training on JSIR raises every metric, by 3--6 points in JAAH and 16--25 points in JSIR; initializing from the pop checkpoint adds another 3--5 points in JAAH, but only about one point for JSIR. The gains concentrate on seventh chords, whose tetrad-level accuracy on JSIR roughly doubles, whereas plain triads are now often labeled with an added seventh. Performance grows monotonically with training-set size, measured on a single hold-out split with a quarter, half, and all of its training recordings (lower block of Table~\ref{tab:results}b): a quarter of the training set (54 recordings, 4~h) already matches the pop model on JAAH, and the curve has not flattened at the full size.

\subsection{Structure Analysis}

We trained SongFormer on the six-label taxonomy of JSIR, from scratch and by fine-tuning the released pop-trained checkpoint, with five-fold cross-validation and three seeds. Following evaluation methods in the field~\cite{smith2011salami, nieto2019harmonix, hao2025songformer}, we report boundary hit-rate F-measures at 0.5~s and 3~s (HR.5F, HR3F), the pairwise frame-clustering F-measure (PWF), and frame-level label accuracy. Because JSIR recordings consist of few, long sections (median four interior boundaries), the customary hit rate, which credits the shared start and end, awards 49\% to an empty prediction; we therefore also report trimmed variants over interior boundaries only.

Both models reach 85\% label accuracy. With the released peak picking, however, they predict 2.4--2.8 boundaries per annotated one, splitting long solos at every chorus, and the trimmed boundary F-measure stays below 50\%. A boundary threshold tuned on the validation split halves the number of boundaries and lifts it to 68--70\% at a cost of one to two points of accuracy. Detected boundaries are beat-accurate (median offset 0.08~s), so the 0.5~s and 3~s tolerances nearly coincide. Pop pre-training does not improve labels, and once the threshold is tuned the fine-tuned model is slightly worse on every metric. The remaining errors are jazz-specific: outros are confused with heads, trades with solos, and interludes are rarely detected, whereas changes of soloist are found with 86\% recall. A fifteen-class variant using the instrument tags reaches 79\% accuracy and identifies the soloing instrument reliably; its main confusion is head versus solo of the same instrument.

\subsection{Implications}

Similar observations recur across tasks. Models trained on popular music transfer poorly to jazz, and their errors fall where Section~\ref{sec:related} located them: at the bar level rather than the pulse, on seventh chords rather than roots, and on section density rather than section labels. A modest amount of in-domain data addresses these deficits: 20--24 hours of JSIR yield a significant downbeat gain on independent jazz sets with at most a marginal loss elsewhere, raise chord accuracy by up to 25 points, and train a structure model to 85\% label accuracy from scratch, whereas pop pre-training adds little. These trends, as well as the monotonic improvements in chord estimation, suggest that annotated jazz audio, not modeling capacity, is the binding constraint, and that a dataset of JSIR's size already relieves it.

\section{Conclusion and Discussion}
\label{sec:conclusion}

In this paper, we presented JSIR, an audio-aligned dataset of jazz standard recordings containing beat, chord, and structure annotations. We demonstrated its potential in MIR research with baseline experiments in three downstream tasks; JSIR on its own brought about meaningful performance gains in all experiments. The annotations are publicly shared along with a web-based browsing tool.
While we are optimistic about the dataset's comprehensiveness, we also acknowledge its limitations: the problem of annotator subjectivity, especially in chord labels, cannot be entirely eliminated; and per-section labels of active instruments, often provided in other datasets, are missing from JSIR. We hope to address these issues by open-sourcing the annotation tool and allowing outside contributions.

As a main driving force behind the evolution of western music in the 20th century, jazz can bring invaluable insights to MIR and computational musicology; we hope that our work helps bring new attention to the matter.

{
\small
\bibliographystyle{IEEEbib}
\bibliography{refs}
}

\end{document}